\documentclass[
reprint,
superscriptaddress,
showpacs,
preprintnumbers,
nofootinbib,
bibnotes,
amsmath,
amssymb, 
aps,
prl,
notitlepage,
floatfix
]{revtex4-1}

\usepackage[utf8]{inputenc}
\usepackage[normalem]{ulem}
\usepackage{graphicx}
\usepackage{dcolumn}

\usepackage[colorlinks=true,
            allcolors=purple,
            pdftitle={Towards a Robust Exclusion of the Sterile-Neutrino Explanation of Short-Baseline Anomalies},
            pdfauthor={Ohana Benevides Rodrigues et. al.},
            ]{hyperref}
\usepackage{url}
\usepackage{enumerate}

\usepackage{slashed,multirow,relsize,soul,feynmp-auto,tikz}
\usepackage{color}
\usepackage{mathrsfs} 
\usepackage{amsmath}
\usepackage{cancel}
\usepackage{bbold}
 \usepackage{mathrsfs}
\usepackage{braket}
\usepackage{physics}
\usepackage{multirow}
\usepackage[capitalize]{cleveref}
\usepackage{xspace}

\usepackage{fontawesome} 
\definecolor{nicegreen}{rgb}{0., 0.75, 0.46}

\definecolor{MH}{rgb}{0.0,0.6,9}
\definecolor{palatinate}{rgb}{0.494, 0.192, 0.482}
\definecolor{blue-violet}{rgb}{0.33, 0.17, 0.89}

\renewcommand{\phi}{\varphi}

\newcommand{\mb}{MB\xspace}
\newcommand{\ub}{$\mu$B\xspace}

\begin{document}
\preprint{FERMILAB-PUB-25-0098-T, MI-HET-852}
\title{Towards a Robust Exclusion of the\\ Sterile-Neutrino Explanation of Short-Baseline Anomalies}

\author{Ohana Benevides Rodrigues}
\email{obenevidesrodrigues@iit.edu}
\affiliation{Department of Physics, Illinois Institute of Technology, Chicago, Illinois 60616}
\author{Matheus Hostert}
\email{mhostert@g.harvard.edu}
\affiliation{Department of Physics \& Laboratory for Particle Physics and Cosmology, Harvard University, Cambridge, MA 02138, USA}
\author{Kevin J. Kelly}
\email{kjkelly@tamu.edu}
\affiliation{Department of Physics and Astronomy, Mitchell Institute for Fundamental Physics and Astronomy, Texas A\&M University, College Station, TX 77843, USA}
\author{Bryce Littlejohn}
\email{blittlej@iit.edu}
\affiliation{Department of Physics, Illinois Institute of Technology, Chicago, Illinois 60616}
\author{Pedro A.~N.~Machado}
\email{pmachado@fnal.gov}
\affiliation{Theoretical Physics Department, Fermilab, P.O. Box 500, Batavia, IL 60510, USA}
\author{Ibrahim Safa}
\email{isafa94@gmail.com}
\affiliation{Columbia University, New York, NY, 10027, USA}
\author{Tao Zhou}
\email{taozhou@tamu.edu}
\affiliation{Department of Physics and Astronomy, Mitchell Institute for Fundamental Physics and Astronomy, Texas A\&M University, College Station, TX 77843, USA}

\date{\today}

\begin{abstract}

The sterile neutrino interpretation of the LSND and MiniBooNE neutrino anomalies is currently being tested at three Liquid Argon detectors: MicroBooNE, SBND, and ICARUS.
It has been argued that a degeneracy between $\nu_\mu \to \nu_e$ and $\nu_e \to \nu_e$ oscillations significantly degrades their sensitivity to sterile neutrinos.
Through an independent study, we show two methods to eliminate this concern. 
First, we resolve this degeneracy by including external constraints on $\nu_e$ disappearance from the PROSPECT reactor experiment.
Second, 
by properly analyzing the full three-dimensional parameter space, we demonstrate that the stronger-than-sensitivity exclusion from MicroBooNE alone already covers the entire 2$\sigma$ preferred regions of MiniBooNE at the level of $2-3\sigma$.
We show that upcoming searches at SBND and ICARUS can improve on this beyond the $4\sigma$ level, thereby providing a rigorous test of short-baseline anomalies.
\end{abstract}

\maketitle

\paragraph{\textbf{Introduction --}} For decades, measurements of neutrino interactions have presented unexpected results, leading to new interpretations of the Standard Model (SM) of particle physics and beyond. In one such case, the MiniBooNE experiment (heretofore, \mb) observed a significant excess ($4.8\sigma$ above backgrounds) of electron-neutrino-like events at low energies~\cite{MiniBooNE:2020pnu}, still to be resolved.
This excess, which has thus far not been explained by conventional neutrino-nucleus interactions or unknown backgrounds~\cite{Hill:2009ek, Hill:2010zy, Brdar:2021ysi, Kelly:2022uaa}, has prompted extensive theoretical and experimental investigations~\cite{Acero:2022wqg} and bears significant implications for particle physics and cosmology. 

The simplest interpretation of this anomaly involves eV-scale sterile neutrinos -- SM-singlet fermions that can mix with active neutrinos driving short-baseline oscillations. 
This interpretation is also historically what connected MiniBooNE to the anomalous results of the Liquid Scintillator Neutrino Detector (LSND) experiment~\cite{LSND:2001aii}, a $3.8\sigma$ excess of $\bar{\nu}_e$ events in a neutrino source from $\mu^+$ decays at rest.
While global fits involving dozens of experiments have established a tension between these anomalies and other results~\cite{Dentler:2018sju, Diaz:2019fwt, Boser:2019rta}, direct tests of the MiniBooNE anomaly using experiments with similar neutrino beams and energy ranges can provide particularly valuable insights into the nature of the excess.

The MicroBooNE experiment (\ub) was specifically designed to test the \mb~anomaly using the superior particle identification capabilities of liquid argon time projection chamber (LArTPC) technology~\cite{MicroBooNE:2015bmn}.
Operating in the same neutrino beam as \mb, the Booster Neutrino Beam, \ub~can distinguish between electron-like events and other backgrounds such as single photons and $\pi^0$ with unprecedented precision~\cite{MicroBooNE:2021zai,MicroBooNE:2021tya,MicroBooNE:2021nxr,MicroBooNE:2021pvo,MicroBooNE:2021wad}.
This enhanced background discrimination significantly reduces systematic uncertainties associated with photon-related backgrounds, which were an important source of uncertainty in the \mb~analysis~\cite{Hill:2010zy,Wang:2014nat}.

Recent sterile-neutrino analyses have supported the idea that \ub~data remain compatible with \mb's allowed region~\cite{Arguelles:2021meu,Denton:2021czb,MicroBooNE:2022sdp}.
Three sterile-neutrino parameters govern oscillations relevant to \mb~and~\ub: two mixing angles and a mass-squared splitting $\Delta m^2_{41}$. 
The mixing elements $U_{e4}$ and $U_{\mu4}$ individually control $\nu_e$ and $\nu_\mu$ disappearance, while their product determines the appearance probability through $\sin^2(2\theta_{\mu e}) \equiv 4|U_{e4}|^2|U_{\mu4}|^2$. 
The copious $\nu_\mu$ events are often used as a control sample to mitigate flux and cross section uncertainties.
However, the Booster Neutrino Beam contains an intrinsic  component of $\nu_e$, at the level of $0.5\%$, representing a significant background to appearance searches.
This background introduces an inescapable degeneracy between $\nu_\mu \to \nu_e$ appearance and $\nu_e \to \nu_e$ disappearance signals for some combination of mixing parameters.

We show here that when profiling over a multi-dimensional parameter space and showing allowed regions in $\sin^2(2\theta_{\mu e})$ versus $\Delta m^2_{41}$, comparisons between different experiments can be misleading.
Difficulties arise from the fact that, for a common $\sin^2(2\theta_{\mu e})$ point in this parameter space, two experiments' underlying best-fit parameters $|U_{e4}|^2$ and $|U_{\mu 4}|^2$ need not be the same.
By examining the full three-dimensional parameter space ($|U_{e4}|^2$, $|U_{\mu4}|^2$, $\Delta m^2_{41}$), we reveal stronger tensions between \mb~and \ub~than previously recognized. 
Our analysis accounts for how these parameters influence the control sample and background predictions in both \mb~and \ub. 
We demonstrate that \ub~can rule out \mb's entire $2\sigma$ preferred region at more than $95\%$ confidence level when properly accounting for all parameter correlations.  
We also examine how sterile-neutrino exclusion is improved when considering electron-flavor disappearance data from the reactor-based PROSPECT experiment constraints and future data from the Fermilab SBN program -- both in terms of profiled $\sin^2(2\theta_{\mu e})$-$\Delta m^2_{41}$ space and the un-profiled full three-dimensional parameter space.

\begin{figure}[t]
    \centering
    \includegraphics[width=\linewidth]{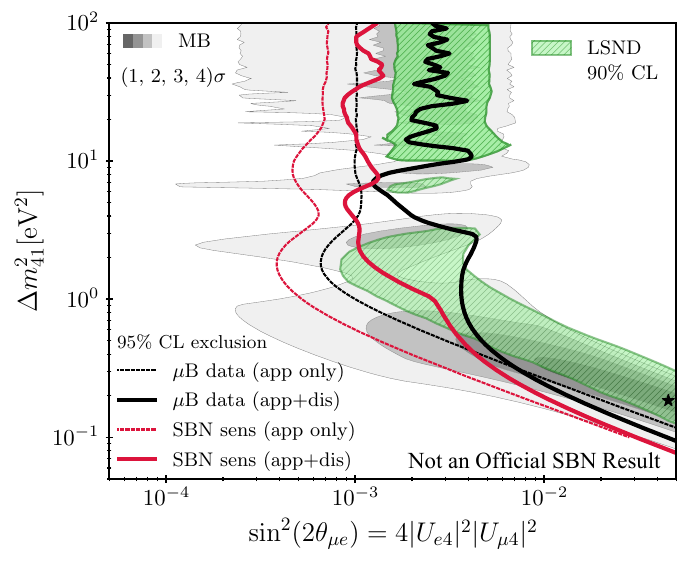}
    \caption{
    The sterile neutrino parameter space and $1,2,3,4\sigma$ preferred regions of MiniBooNE (grey regions) and $90\%$ CL region of LSND (green region) along with the $95\%$ CL MicroBooNE exclusion (black lines) and SBN sensitivity (red lines), assuming Wilks' theorem.
    Solid lines show the result of a full $3+1$ oscillation model and dashed lines of the common but physically-inconsistent, appearance-only model.
    The MiniBooNE fit is shown for the full oscillation model.
    For LSND, they are approximately the same.
    The apparent compatibility between  experiments is a potential artifact of the projection into the appearance mixing angle.
    }
    \label{fig:app_only}
\end{figure}

\paragraph{\textbf{Short-baseline Oscillations \& Degeneracies --}}
The sterile neutrino interpretation of the \mb~anomaly requires extending the standard three-neutrino framework to include a fourth mass eigenstate. 
At short baselines, the oscillation probabilities in this framework take a simple form. 
The appearance probability is given by
\begin{equation}
    P(\nu_\mu \to \nu_e) =  \sin^2(2\theta_{\mu e})\sin^2\left(\frac{\Delta m^2_{41} L}{4E}\right),
\end{equation}
while for disappearance
\begin{equation}
    P(\nu_\alpha \to \nu_\alpha) = 1 - \sin^2(2\theta_{\alpha\alpha})\sin^2\left(\frac{\Delta m^2_{41} L}{4E}\right),
\end{equation}
where $\sin^2(2\theta_{\alpha\alpha})\equiv 4|U_{\alpha4}|^2(1-|U_{\alpha4}|^2)$ for $\alpha=e,\mu$.

In short-baseline oscillation experiments, such as \mb~and \ub, intrinsic $\nu_e$ backgrounds are affected by the same mixing parameters that control appearance. 
The $\nu_e$ disappearance probability scales with $U_{e4}$, partially compensating any appearance signal. 
Additionally, the $\nu_\mu$ flux used to predict the appearance signal is inferred from $\nu_\mu$ measurements, which themselves can be depleted by oscillations from nonzero $U_{\mu4}$. 
This interplay between appearance, disappearance, and flux determination introduces degeneracies with (potentially) different manifestations in each experiment's parameter space.

The detector technologies employed by \mb~and \ub~lead to different background compositions. 
\mb's Cherenkov detector cannot distinguish electrons from single photons that convert to $e^+e^-$ pairs inside the fiducial volume, making $\pi^0$ decays and $\Delta(1232)\to\gamma N$ processes significant backgrounds. 
The LArTPC technology of \ub\ allows to distinguish electrons from photons using the presence of a gap between the event vertex and the electromagnetic shower, as well as the ionization yield in the first few centimeters of the shower, substantially reducing photon-related backgrounds.
These different background compositions translate into different degeneracies in the sterile neutrino parameter space, which is crucial to estimate the compatibility between these two experiments, as we will see shortly.

\paragraph{\textbf{Datasets and Simulations --}}
We analyze MiniBooNE and MicroBooNE data using their data releases~\cite{MiniBooNE:2021bgc, MicroBooNE:2021nxr}, treating backgrounds consistently within the 3+1 oscillation framework. 
To estimate the sensitivity of the SBN program, we build on our existing \ub~framework~\cite{Arguelles:2021meu,Hostert:2024etd}.
Complementary electron neutrino disappearance constraints from the PROSPECT reactor neutrino experiment are also examined in this study.  
The PROSPECT experiment deployed a 4~ton segmented liquid scintillator detector target a distance of 7-9 meters from the High Flux Isotope Reactor (HFIR) at Oak Ridge National Laboratory~\cite{PROSPECT:2018dnc}, detecting over 60,000 inverse beta decay interactions ($\overline{\nu}_e + p \rightarrow e^+ + n$) while searching for short-baseline $\overline{\nu}_e$ disappearance~\cite{PROSPECT:2020sxr,PROSPECT:2022wlf}.  
Rather than re-analyzing PROSPECT neutrino and background datasets, we use maps of $\Delta \chi^2$ versus $\sin^2(2\theta_{ee})$ and $\Delta m^2_{41}$ provided by PROSPECT in their final oscillation analysis using the full dataset from the now-decommissioned PROSPECT-I detector~\cite{PROSPECT:2024gps}.  

\begin{figure}[t]
    \centering
    \includegraphics[width=\linewidth]{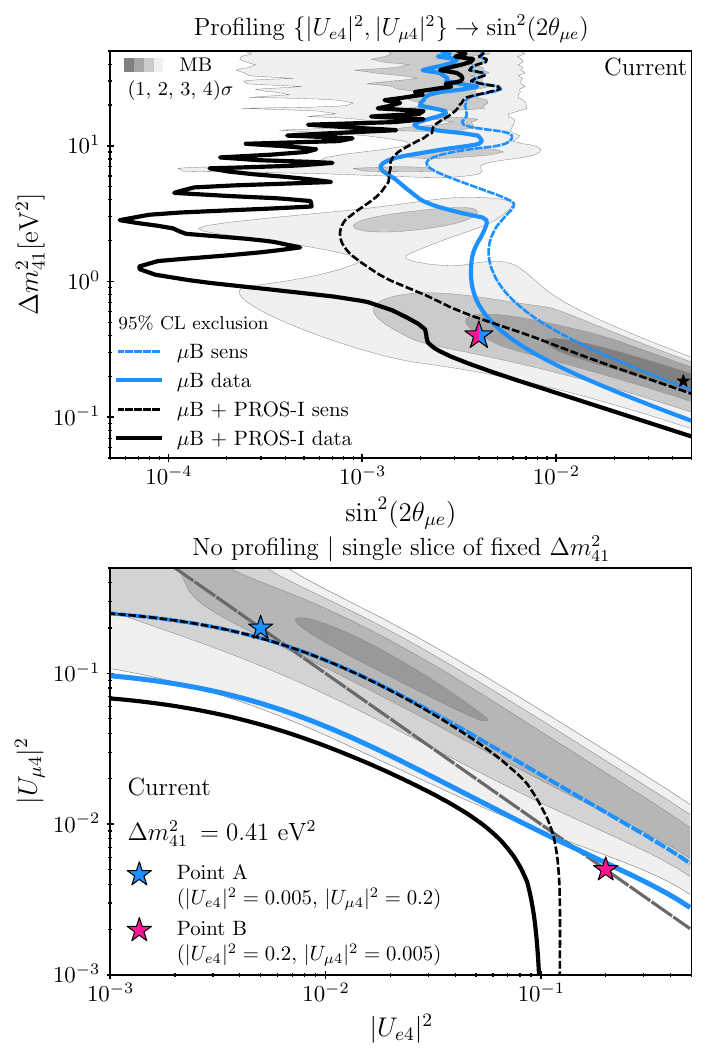}
    \caption{
    Top) Typical sterile-neutrino parameter space presentation, comparing the MiniBooNE-preferred regions (grey), with
    95\% CL sensitivity/exclusion of MicroBooNE in blue (+PROSPECT in black), assuming Wilks' theorem. 
    Exclusions (data) are presented as thick, solid lines; sensitivities with thin, dashed lines.
    \\
    Bottom) Slice of $\Delta m_{41}^2 = 0.41$~eV$^2$ parameter space, allowing $\left\lbrace |U_{e4}|^2,~|U_{\mu 4}|^2\right\rbrace$ to vary independently \textit{before profiling}. Blue (A) and red (B) stars both correspond to $\sin^2(2\theta_{\mu e}) = 4\times 10^{-3}$ in the top panel. Despite this angle not being covered by \ub~in the profiled, top panel, the line it represents in the bottom panel contains no points compatible with \emph{both} \mb~and \ub.
    \label{fig:main_profiled_comparison}
    }
\end{figure}


\paragraph{\textbf{Pitfalls of Projection --}}
Traditional sterile neutrino searches for anomalous $\nu_\mu \to \nu_e$ appearance -- particularly LSND, \mb, and \ub~-- present their results in the $\sin^2(2\theta_{\mu e})$-$\Delta m^2_{41}$ plane, as we show in \cref{fig:app_only}. 
While optimal for experiments that are sensitive almost exclusively to $\nu_\mu \to \nu_e$ appearance signals like LSND, this projection obscures important physical effects in experiments with significant neutrino-related backgrounds like \mb\ and \ub.
The latter experiments are sensitive not just to the product  $4|U_{e4}|^2|U_{\mu 4}|^2$ via the appearance channel, but to individual mixing elements through $\nu_\mu$ disappearance in flux-normalizing control samples and $\nu_e$ disappearance in backgrounds.
Due to this scaling, a small appearance signal is typically accompanied by much larger disappearance effects.
This leads to a nontrivial degeneracy among the mixing elements and the appearance amplitude.

\cref{fig:app_only} illustrates the impact of the above degeneracy: both \ub\ and SBN sensitivities assuming the unphysical scenario of appearance-only oscillations 
are significantly stronger than in the full 3+1 model, 
which accounts for oscillations in the backgrounds and control sample.
Within the consistent treatment of oscillations, this result na\"ively suggests that there currently is a large parameter space where concordance between \mb and \ub can be found and that SBN may not even be able to cover the entire $3\sigma$ region of the \mb parameter space at more than $95\%$ CL.
However, as we show below, this conclusion is incorrect.



One avenue to reduce the impact of the degeneracy is to include data from a neutrino source with a different composition in terms of its $\nu_e$/$\nu_\mu$ flavor ratio.  
This is present in the NuMI beam and is currently being pursued by the \ub collaboration~\cite{MicroBooNE-NOTE-1129-PUB,MicroBooNE-NOTE-1132-PUB}.
Indeed, the measurement of this flux at MiniBooNE~\cite{MiniBooNE:2008hnl} has already been used to constrain oscillations in Ref.~\cite{Diaz:2019fwt}, although, by itself, it is not very sensitive to sterile neutrinos.
With a larger $\nu_e$/$\nu_\mu$ flavor ratio, appearance and disappearance signals can be directly constrained, and parameters for which $\nu_e$ disappearance can accommodate a greater amount of $\nu_\mu \to \nu_e$ appearance in the BNB flux without leading to an observable excess would not necessarily do so for the NuMI dataset, thereby reducing the impact of the degeneracy shown in \cref{fig:app_only}. 

The flavor-pure flux of $\overline{\nu}_e$ emanating from a nuclear reactor also offers an opportunity for reducing  degeneracy in oscillation results from mixed-flavor neutrino beams.  
In \cref{fig:main_profiled_comparison}(top) we present the \ub~exclusion in this parameter space with and without PROSPECT data.  
PROSPECT provides invaluable direct constraints on $|U_{e4}|^2$ through searches for short-baseline $\bar{\nu}_e$ disappearance. 
The combination of appearance and disappearance experiments naturally resolves the degeneracy in the 3+1 framework by directly imposing a limit on the amount of $\nu_e$ disappearance that is permissible in the intrinsic background, effectively removing the possibility to hide appearance signals.
Combined with PROSPECT, the BNB experimental searches recover their constraining power in the traditional profiled parameter space of \cref{fig:app_only,fig:main_profiled_comparison}.


We note that the external PROSPECT constraints accentuates the difference between \ub's sensitivity and exclusion: the constraint on $\sin^2(2\theta_{\mu e})$ for some values of $\Delta m_{41}^2$ is almost an order of magnitude stronger than the expected sensitivity.
This is due to three contributing factors.
First, the observed excess of $\nu_\mu$ events at \ub disfavors $\nu_\mu$ disappearance.
Second, and more importantly, the accompanying observed deficit of $\nu_e$ events leaves less room for $\nu_\mu \to \nu_e$ appearance.
Finally, upon accounting for external constraints on $\nu_e$ disappearance from PROSPECT, any amount of $\nu_\mu \to \nu_e$ appearance is strongly disfavored.

\begin{figure}[t]
    \centering
    \includegraphics[width=\linewidth]{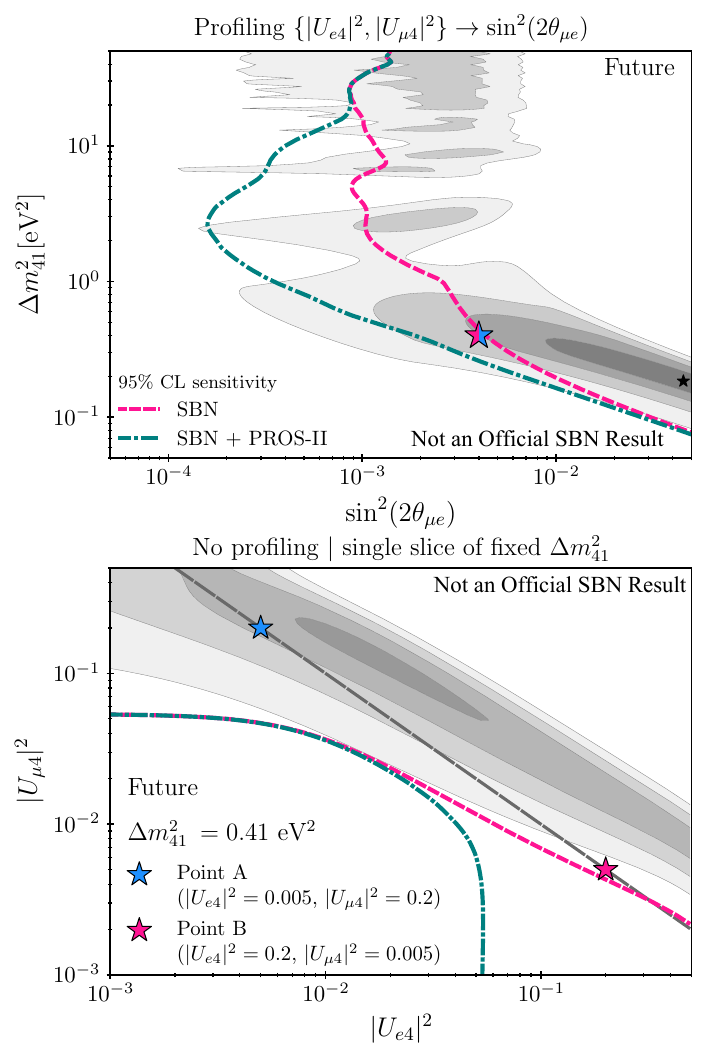}
    \caption{Same as~\cref{fig:main_profiled_comparison}, focusing on future projections of SBN (pink) and SBN+PROSPECT-II (green) capabilities.
    \label{fig:main_profiled_comparison_2}
    }
\end{figure}

Apart from adding new datasets, another way to confront this degeneracy is to investigate the parameter space without relying on profiling over unseen parameters.  
To illustrate this point, we take two benchmark points with the same appearance angle $\sin^2(2\theta_{\mu e}) = 0.004$ and mass splitting $\Delta m^2_{41} = 0.41$~eV$^2$ but different mixing patterns: (A) $|U_{e4}|^2 = 0.005$ and $|U_{\mu 4}|^2 = 0.2$ and (B) $|U_{e4}|^2 = 0.2$ and $|U_{\mu 4}|^2 = 0.005$.
In the top panel of \cref{fig:main_profiled_comparison}, these two points appear as the half-blue/half-pink star, apparently compatible both with \mb's preference for the existence of a sterile neutrino and compatible with \ub's null result -- according to the profiled parameter space, \ub lacks the statistical power to meaningfully exclude this point.
However, in the fixed-$\Delta m_{41}^2$ slice of the parameter space in the bottom panel, these two points are distinct and neither of them is simultaneously compatible with \emph{both} \mb and \ub.
Indeed, the entire dashed gray line where $\sin^2(2\theta_{\mu e})=0.004$ in the slice panel collapses onto the star of the profiled plane, but nowhere along it can the preference regions of \mb~and \ub be simultaneously satisfied.  


While our benchmark points (A) and (B) yield identical $\sin^2(2\theta_{\mu e})$, they produce distinct event rates in the $\nu_\mu$ sample, and in the $\nu_e$ sample due to different background compositions. 
This is explicitly shown in \cref{fig:event_rates} of the Endmatter, where we compare predicted event spectra for \ub's $\nu_e$ and $\nu_\mu$ samples, as well as \mb electron-like events.  
The relatively large value of $|U_{\mu4}|^2$ of point (A) is disfavored by \ub as it significantly affects the $\nu_\mu$ channel.
Point (B), on the other hand, has a smaller value of $|U_{\mu4}|^2$, barely changing $\nu_\mu$ disappearance, and is allowed by \ub data. 
Nevertheless, the larger value of $|U_{e4}|^2$ leads to too much disappearance of the intrinsic $\nu_e$ background, not providing enough excess of electron-like events in \mb to be favored by the data.

The aforementioned degeneracy is therefore effectively broken, and a fundamental tension between \ub and \mb is revealed in the full three-dimensional parameter space of sterile neutrinos that would otherwise remain obscured in traditional profiled comparisons. 
This point is demonstrated by \cref{fig:main_profiled_comparison}(bottom) for a specific choice of $\Delta m^2_{41}$, and is generalized for the broader parameter space in \cref{fig:Slices_2dof} of the Endmatter, where we show more slices of the parameter space for 3 degrees of freedom.  
While the profiled space provides a picture of potential consonance between \mb and \ub at the chosen $\Delta m^2_{41}$, the full space in \cref{fig:main_profiled_comparison} shows no overlap between the \ub 95\% CL allowed region and \mb's 3$\sigma$ allowed region.  Adding PROSPECT excludes all of \mb's 4$\sigma$ allowed region at this $\Delta m^2_{41}$.
Across the full phase space, we observe that no single combination of mixing parameters in the 3+1 scenario can simultaneously explain \mb's excess at $2\sigma$ while remaining consistent with \ub's observations at the same confidence level.  


\paragraph{\textbf{Future Projections --}} 
With future data from SBND and ICARUS, the capability of excluding the three-dimensional \mb-preferred parameter space is even stronger.
To simulate these, we make the simplified assumption that SBND and ICARUS will have the same signal efficiency as \ub, as well as the same number of backgrounds per neutrino per fiducial mass.
This is a conservative estimate, as SBND and ICARUS are expected to improve upon detector performance, event reconstruction, and analysis techniques \cite{ICARUS:2023gpo, SBND:2024vgn}.
With our procedure, SBND and ICARUS share the same energy binning and resolution as the \ub~analysis discussed above.
We take the fiducial mass to be 112~t and 476~t of liquid argon for SBND and ICARUS, respectively, with a total exposure of $6.6 \times 10^{20}$ POT. 
The neutrino flux at the location of SBND and ICARUS is rescaled from the \ub~flux by an overall energy-independent factor of $\left(L_{\mu{\rm B}}/L_{\rm SBND/ICARUS}\right)^2$, where $L_{\rm SBND} = 110$~m, $L_{\rm ICARUS} = 600$~m, and $L_{\mu{\rm B}} = 470$~m are the approximate detector distances to the target.
More details can be found in the Endmatter.  
Projecting forward, we also consider future reactor antineutrino observations from PROSPECT-II. 
We use $\Delta \chi^2$ maps provided by the collaboration that describe the expected sensitivity of a two-year run at HFIR with an upgraded detector~\cite{PROSPECT:2021jey}.  

In~\cref{fig:main_profiled_comparison_2}, we present the same comparison of \mb-preferred parameter space against future projections including SBN 
and additionally PROSPECT-II. 
In the top panel, the (misleading) profiling of 
$\left\lbrace \left\lvert U_{e4}\right\rvert^2,\left\lvert U_{\mu 4}\right\rvert^2\right\rbrace \to \sin^2\left( 2\theta_{\mu e}\right)$ portrays sensitivity that is not as impressive as one would hope -- the SBN-only and SBN + PROSPECT-II projections only demonstrate modest improvement on the current \ub and \ub + PROSPECT results.\footnote{This may, however, be driven by the fluctuations present in \ub~data relative to its expectations.} In contrast, when we focus on the sliced parameter space (bottom panel of~\cref{fig:main_profiled_comparison_2} along with projections in~\cref{fig:Slices_2dof} of the Endmatter), the future improvements are readily apparent. Across the three-dimensional parameter space $\left\lbrace |U_{e4}|^2, |U_{\mu 4}|^2, \Delta m_{41}^2\right\rbrace$, SBN and PROSPECT-II will thoroughly explore the \mb-preferred parameter space.

\begin{figure}[t]
    \centering
    \includegraphics[width=\linewidth]{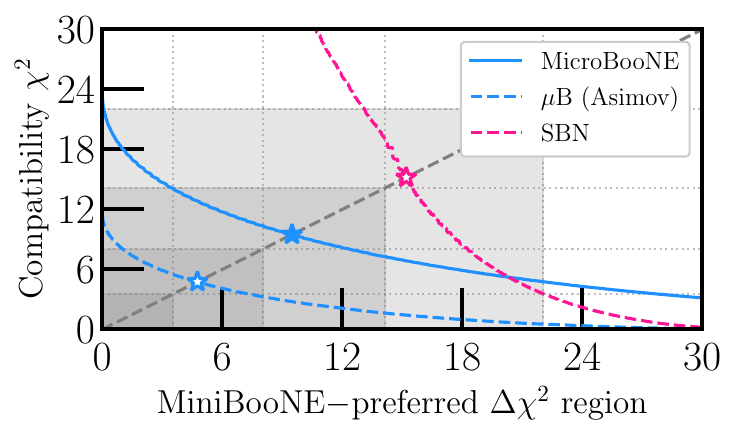}
    \caption{
    The compatibility $\chi^2$ of MicroBooNE (blue) and SBN (pink) compared against the MiniBooNE-preferred parameter space, as described in the text. Each experiments' intersection with the diagonal grey line is of particular emphasis.
    Grey regions contain the regions of compatibility, assuming Wilks' theorem, starting at $1\sigma$ (darkest grey) to $4\sigma$ (lightest grey).
    \label{fig:CompatibilityChiSq}
    }
\end{figure}

\paragraph{\textbf{Compatibility with MiniBooNE --}}
To quantify the compatibility between any given null results and the \mb-preferred multi-dimensional space, we construct a compatibility $\chi^2$.
For a given value $\Delta \chi^2_0$, we obtain the 3D region in the sterile neutrino parameter space satisfying $\Delta \chi^2_{\rm MB} < \Delta \chi^2_{0}$.
Then, scanning that entire region, we quantify the minimum $\Delta \chi^2_{\rm exp}$ of another experiment within that region -- determining the level at which that experiment excludes the entirety of the \mb-preferred $\Delta \chi^2$ region.\footnote{
More precisely, given a region of the model parameter space defined by $\vec\vartheta\left(\Delta \chi^2_0\right) = \left\lbrace \vartheta | \Delta \chi^2_{\rm MB} < \Delta \chi^2_0\right\rbrace$, the compatibility $\chi^2$ for a given experiment or set of experiments is defined as $\chi^2_{\rm exp}\left( \Delta \chi^2_0\right) = \mathrm{min}_{\vec\vartheta(\Delta\chi^2_0)} \left \lbrace \Delta \chi^2_{\rm exp}\right \rbrace$.
}
This is presented for \ub and its Asimov sensitivity expectation, and for the projection for SBN in~\cref{fig:CompatibilityChiSq}.


Each curve's intersection with the dashed grey diagonal line addresses the question: ``at what confidence level $X$ can this experiment exclude \mb's preference at the same confidence level?'' This corresponds to $\Delta\chi^2=9.5$ for \ub~alone (compared to $\Delta\chi^2=4.8$ for its sensitivity), which increases modestly by adding PROSPECT. 
With three degrees of freedom and assuming Wilks' theorem, this means \ub excludes \mb's entire $2.3\sigma$ ($97.6\%$~CL) region at greater than the same confidence level.
This should be contrasted with the compatibility calculated using the \ub sensitivity, which yields $1.3\sigma$ ($80.9\%$ CL).
This quantitative analysis confirms our parameter space exploration -- no point in the full three-dimensional parameter space can simultaneously explain \mb's excess while remaining consistent with \ub's observations at more than $2\sigma$.\footnote{The addition of PROSPECT data for improving this exclusion is only modest: \ub + PROSPECT excludes \mb's entire $2.4\sigma$ ($98.5$\% CL) at greater than the same confidence level. In contrast, the Asimov-expected exclusion of \ub + PROSPECT was $1.7\sigma$ ($91.3\%$ CL).}
We also see that SBN will improve on this substantially, covering the entire $3\sigma$ region of \mb at more than $3\sigma$ while excluding MB's most-favored ${\sim}2\sigma$ parameter space with extremely high ($>$5$\sigma$) confidence.  

\paragraph{\textbf{Discussion -- }}
We have presented two complementary arguments that show that the MicroBooNE experiment and the SBN program will robustly cover the sterile neutrino interpretation of the LSND and MiniBooNE anomalies.
This is the first time that the MicroBooNE exclusion and the SBN sensitivities are shown throughout the full $3+1$-oscillation model without resorting to the statistical procedure of profiling.  
In this full oscillation model, the coexistence of a $\nu_\mu \to \nu_e$ appearance signal and the disappearance of $\nu_{e}$ backgrounds and $\nu_\mu$ in the control sample leads to a degeneracy and an apparent loss of experimental sensitivity to the model in the plane of $\sin^2(2\theta_{\mu e})$ vs $\Delta m^2_{41}$, after profiling.  
We conclude, however, that when comparing allowed experimental regions \textit{before} profiling, this effect does not substantially diminish the exclusion power of \ub data -- we were able to achieve this by generalizing the MiniBooNE, MicroBooNE, and SBN fits to slices of the full three-dimensional parameter space: $(|U_{e4}|^2, |U_{\mu 4}|^2, \Delta m^2_{41})$.
Exploring the full three-dimensional parameter space provides qualitatively different conclusions.  
For this reason, we urge all experiments that are sensitive to multiple oscillation parameters to make their full multi-dimensional $\chi^2$ maps available, including MiniBooNE, to enable these compatibility comparisons in the future by the SBN program.  

Even in the profiled case, we show that the degeneracy in \ub data can be avoided altogether by incorporating strong existing  constraints on $\nu_e$ disappearance from the flavor-pure PROSPECT reactor neutrino dataset.  
Besides PROSPECT, constraints on electron-flavor disappearance from \ub's NuMI beam datasets~\cite{MicroBooNE-NOTE-1129-PUB,MicroBooNE-NOTE-1132-PUB} and from solar neutrino~\cite{Goldhagen:2021kxe} and absolute reactor neutrino flux measurements~\cite{DayaBay:2014fct,Berryman:2020agd,Giunti:2021kab}, while not examined here, can play a complementary degeneracy-reducing role.
When examining the full three-dimensional oscillation phase space, PROSPECT data unsurprisingly serves as the strongest constraint on large $|U_{e4}|$ values.


Finally, we construct a quantitative measure of the statistical compatibility between any combination of experiments, and illustrate it for MiniBooNE, MicroBooNE, PROSPECT, and SBN in the sterile neutrino model.
By requiring no overlap between the region of preference of MiniBooNE with that of MicroBooNE, we conclude that under this metric, the compatibility between these two experimental results is at the level of $\chi^2 \sim 10$, corresponding to just above $2\sigma$ statistical incompatibility.
Generalizing this result to SBN, we find that eventually the SBN program could drive this compatibility metric to the level of $\chi^2 = 16$, or just above $3\sigma$, while excluding \mb’s most-favored parameter space with extremely high ($>$5$\sigma$) confidence.


\paragraph{Acknowledgements:} 
We thank Joachim Kopp for comments on this manuscript.
The work of MH is supported by the Neutrino Theory Network Program Grant \#DE-AC02-07CHI11359 and the US DOE Award \#DE-SC0020250. KJK and TZ are supported in part by US DOE Award \#DE-SC0010813. OBR and BRL are supported by US DOE Award \#DE-SC0008347. The work of IS was supported by the National Science Foundation Award PHY-2310080. 
This manuscript has been authored by Fermi Forward Discovery Group, LLC under Contract No. 89243024CSC000002 with the U.S. Department of Energy, Office of Science, Office of High Energy Physics.

\emph{Note:} This work is not an official MicroBooNE, SBND, or ICARUS result.

\section{ENDMATTER}

\begin{figure}[t!]
    \centering
    \includegraphics[width=\linewidth]{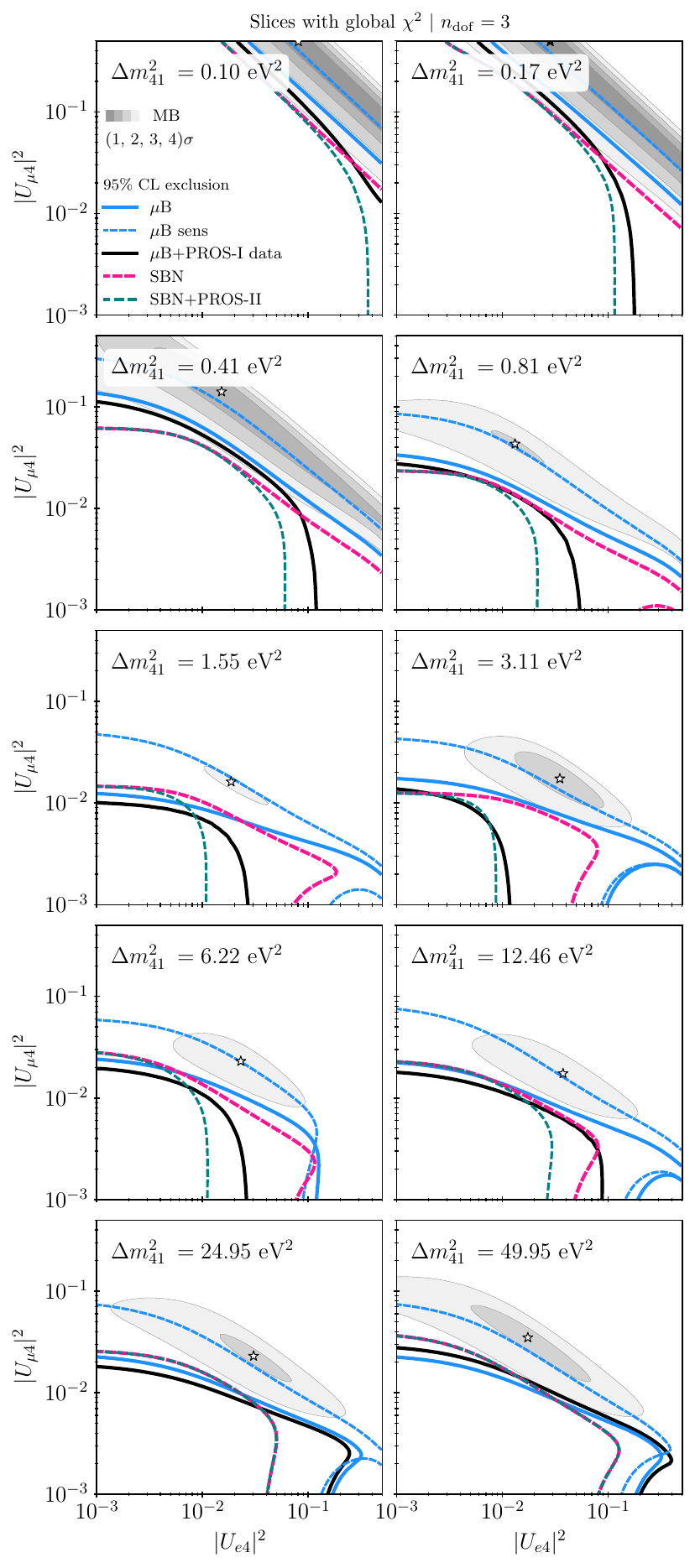}
    \caption{
    The MiniBooNE preferred regions, MicroBooNE exclusion, and SBN sensitivities in slices of fixed $\Delta m^2_{41}$ of the 3-dimensional parameter space of sterile neutrinos for $3$~degrees of freedom (global $\chi^2$).
    Excluded regions are to the top right of each panel.
    }
    \label{fig:Slices_2dof}
\end{figure}

\paragraph{Full slices of parameter space --}
In this appendix we expand on the point made in the main text by showing more slices of the 3-dimensional parameter space of sterile neutrinos.
We compare the MiniBooNE-preferred regions with exclusions from MicroBooNE, PROSPECT, and the sensitivity of SBN with and without PROSPECT-II in \cref{fig:Slices_2dof} using the global $\Delta \chi^2$ with 3 degrees of freedom, and assuming Wilks' theorem.
The result of our coverage test is validated by noting that all MiniBooNE regions are already covered by MicroBooNE at the $95\%$~CL level, even before including PROSPECT data.

\begin{figure}[t!]
    \centering
    \includegraphics[width=0.49\textwidth]{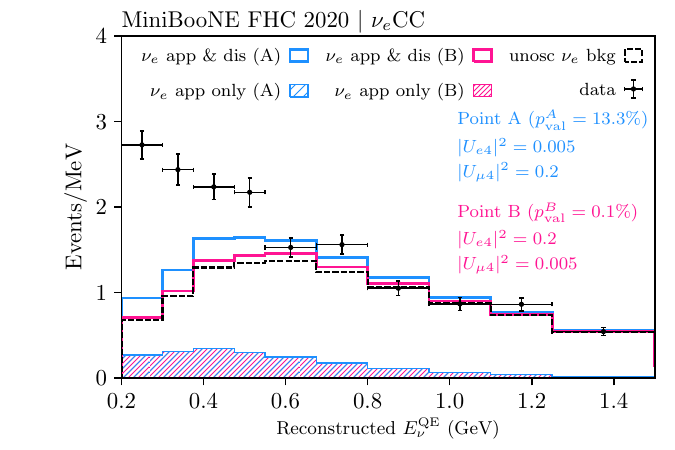}
    \includegraphics[width=0.49\textwidth]{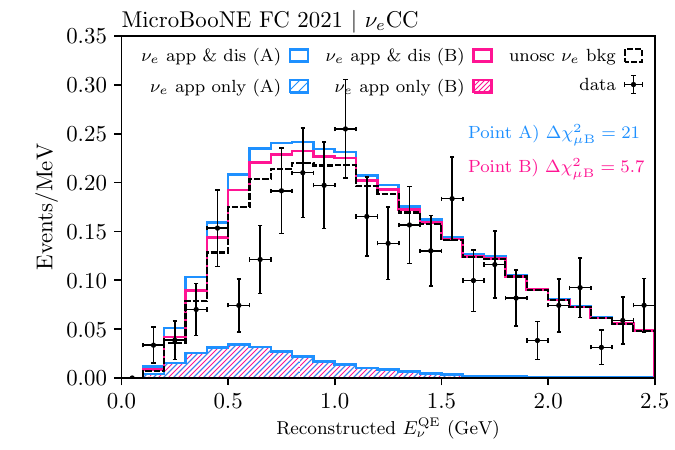}
    \includegraphics[width=0.49\textwidth]{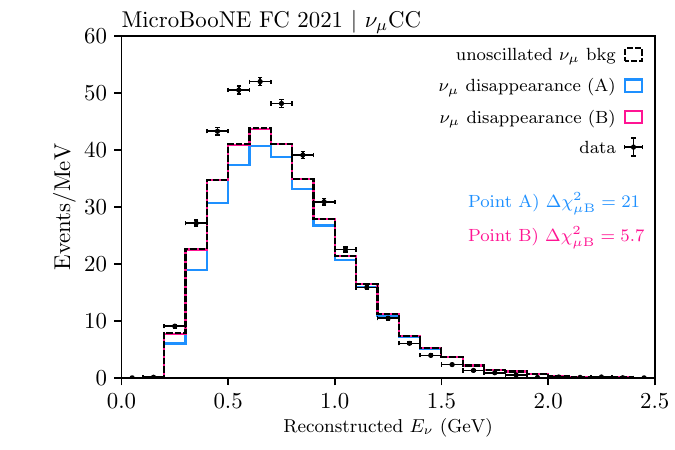}
    \caption{
    The $\nu_e$ event rate at MiniBooNE (top) and MicroBooNE (middle), and $\nu_\mu$ event rate at MicroBooNE (bottom), after subtraction of mis-ID backgrounds.
    We show the theory prediction of a sterile neutrino model for the two benchmark points discussed in \cref{fig:main_profiled_comparison,fig:main_profiled_comparison_2}, both with the same mass splitting $\Delta m^2_{41} = 0.41$~eV$^2$.
    Point A shows a points excluded by \ub~at the $\Delta \chi^2_{\mu B} = 21$ level, and point B shows a point with the $\nu_e$ appearance/disappearance degeneracy that is compatible with \ub at $\Delta \chi^2_{\mu B} = 5.7$.
\label{fig:event_rates}
    }
\end{figure}

In \cref{fig:event_rates}, we also show the event rates for MiniBooNE (top) and MicroBooNE $\nu_e$ sample (middle) and $\nu_\mu$ sample (bottom), after subtraction of mis-ID backgrounds.
For reference, we present the theory predictions of a sterile neutrino model for the two benchmark points shown in \cref{fig:main_profiled_comparison,fig:main_profiled_comparison_2}.

\paragraph{SBN sensitivities --} To estimate the Asimov sensitivity of SBN, we use a $\chi^2$ test statistic and model the systematic uncertainty on the BNB flux and neutrino-Argon cross sections via an approximate covariance matrix.
We analyze the $\nu_\mu$ and $\nu_e$ spectra of SBND and ICARUS together.
We account for energy-dependent cross section uncertainties $\sigma^\text{xs}_i=15\%$, fully correlated between $\nu_\mu$ and $\nu_e$ and between SBND and ICARUS, but uncorrelated over all energies $E_i$.
For flux uncertainties, we take an the $\nu_e$ and $\nu_\mu$  fluxes to be independent.
We consider $\sigma^{e}_i=10\%$, correlated among $\nu_e$ samples in both experiments but uncorrelated over all energies $E_i$, and similar for $\nu_\mu$.

\begin{figure}[t]
    \centering
    \includegraphics[width=\linewidth]{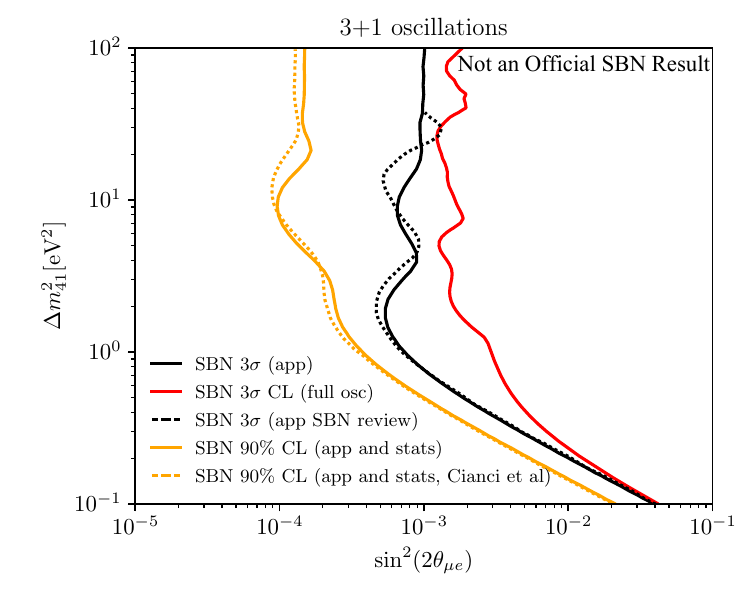}
    \caption{The comparison of our SBN sensitivities with the literature for the case of a sterile model with $\nu_\mu \to \nu_e$ appearance only.
    We compare our sensivity using statistical uncertainty only (solid orange) with that in Ref.~\cite{Cianci:2017okw} (dashed orange).
    We also compare the sensitivity including systematics (solid black) with that in Ref.~\cite{Machado:2019oxb}.
    Also shown is our sensitivity for a full oscillation model (solid red).
    \label{fig:comparison}}
\end{figure}

For validation, we also show the 2-dimensional parameter space of $\Delta m_{41}^2$ vs $\sin^2(2\theta_{\mu e})$ in \cref{fig:comparison}, comparing our result for the simpler, though unphysical, appearance-only model with that of Ref.~\cite{Cianci:2017okw} for the case of statistical uncertainties only, and with the one in Ref.~\cite{Machado:2019oxb} for the appearance-only full systematics case.
No previous SBN sensitivity curve could be found for the case of full oscillations.

\bibliographystyle{apsrev4-1}
\bibliography{main}{}

\begin{thebibliography}{38}%
\makeatletter
\providecommand \@ifxundefined [1]{%
 \@ifx{#1\undefined}
}%
\providecommand \@ifnum [1]{%
 \ifnum #1\expandafter \@firstoftwo
 \else \expandafter \@secondoftwo
 \fi
}%
\providecommand \@ifx [1]{%
 \ifx #1\expandafter \@firstoftwo
 \else \expandafter \@secondoftwo
 \fi
}%
\providecommand \natexlab [1]{#1}%
\providecommand \enquote  [1]{``#1''}%
\providecommand \bibnamefont  [1]{#1}%
\providecommand \bibfnamefont [1]{#1}%
\providecommand \citenamefont [1]{#1}%
\providecommand \href@noop [0]{\@secondoftwo}%
\providecommand \href [0]{\begingroup \@sanitize@url \@href}%
\providecommand \@href[1]{\@@startlink{#1}\@@href}%
\providecommand \@@href[1]{\endgroup#1\@@endlink}%
\providecommand \@sanitize@url [0]{\catcode `\\12\catcode `\$12\catcode `\&12\catcode `\#12\catcode `\^12\catcode `\_12\catcode `\%12\relax}%
\providecommand \@@startlink[1]{}%
\providecommand \@@endlink[0]{}%
\providecommand \url  [0]{\begingroup\@sanitize@url \@url }%
\providecommand \@url [1]{\endgroup\@href {#1}{\urlprefix }}%
\providecommand \urlprefix  [0]{URL }%
\providecommand \Eprint [0]{\href }%
\providecommand \doibase [0]{http://dx.doi.org/}%
\providecommand \selectlanguage [0]{\@gobble}%
\providecommand \bibinfo  [0]{\@secondoftwo}%
\providecommand \bibfield  [0]{\@secondoftwo}%
\providecommand \translation [1]{[#1]}%
\providecommand \BibitemOpen [0]{}%
\providecommand \bibitemStop [0]{}%
\providecommand \bibitemNoStop [0]{.\EOS\space}%
\providecommand \EOS [0]{\spacefactor3000\relax}%
\providecommand \BibitemShut  [1]{\csname bibitem#1\endcsname}%
\let\auto@bib@innerbib\@empty
\bibitem [{\citenamefont {Aguilar-Arevalo}\ \emph {et~al.}(2021{\natexlab{a}})\citenamefont {Aguilar-Arevalo} \emph {et~al.}}]{MiniBooNE:2020pnu}%
  \BibitemOpen
  \bibfield  {author} {\bibinfo {author} {\bibfnamefont {A.~A.}\ \bibnamefont {Aguilar-Arevalo}} \emph {et~al.} (\bibinfo {collaboration} {MiniBooNE}),\ }\href {\doibase 10.1103/PhysRevD.103.052002} {\bibfield  {journal} {\bibinfo  {journal} {Phys. Rev. D}\ }\textbf {\bibinfo {volume} {103}},\ \bibinfo {pages} {052002} (\bibinfo {year} {2021}{\natexlab{a}})},\ \Eprint {http://arxiv.org/abs/2006.16883} {arXiv:2006.16883 [hep-ex]} \BibitemShut {NoStop}%
\bibitem [{\citenamefont {Hill}(2010)}]{Hill:2009ek}%
  \BibitemOpen
  \bibfield  {author} {\bibinfo {author} {\bibfnamefont {R.~J.}\ \bibnamefont {Hill}},\ }\href {\doibase 10.1103/PhysRevD.81.013008} {\bibfield  {journal} {\bibinfo  {journal} {Phys. Rev. D}\ }\textbf {\bibinfo {volume} {81}},\ \bibinfo {pages} {013008} (\bibinfo {year} {2010})},\ \Eprint {http://arxiv.org/abs/0905.0291} {arXiv:0905.0291 [hep-ph]} \BibitemShut {NoStop}%
\bibitem [{\citenamefont {Hill}(2011)}]{Hill:2010zy}%
  \BibitemOpen
  \bibfield  {author} {\bibinfo {author} {\bibfnamefont {R.~J.}\ \bibnamefont {Hill}},\ }\href {\doibase 10.1103/PhysRevD.84.017501} {\bibfield  {journal} {\bibinfo  {journal} {Phys. Rev. D}\ }\textbf {\bibinfo {volume} {84}},\ \bibinfo {pages} {017501} (\bibinfo {year} {2011})},\ \Eprint {http://arxiv.org/abs/1002.4215} {arXiv:1002.4215 [hep-ph]} \BibitemShut {NoStop}%
\bibitem [{\citenamefont {Brdar}\ and\ \citenamefont {Kopp}(2022)}]{Brdar:2021ysi}%
  \BibitemOpen
  \bibfield  {author} {\bibinfo {author} {\bibfnamefont {V.}~\bibnamefont {Brdar}}\ and\ \bibinfo {author} {\bibfnamefont {J.}~\bibnamefont {Kopp}},\ }\href {\doibase 10.1103/PhysRevD.105.115024} {\bibfield  {journal} {\bibinfo  {journal} {Phys. Rev. D}\ }\textbf {\bibinfo {volume} {105}},\ \bibinfo {pages} {115024} (\bibinfo {year} {2022})},\ \Eprint {http://arxiv.org/abs/2109.08157} {arXiv:2109.08157 [hep-ph]} \BibitemShut {NoStop}%
\bibitem [{\citenamefont {Kelly}\ and\ \citenamefont {Kopp}(2023)}]{Kelly:2022uaa}%
  \BibitemOpen
  \bibfield  {author} {\bibinfo {author} {\bibfnamefont {K.~J.}\ \bibnamefont {Kelly}}\ and\ \bibinfo {author} {\bibfnamefont {J.}~\bibnamefont {Kopp}},\ }\href {\doibase 10.1007/JHEP05(2023)113} {\bibfield  {journal} {\bibinfo  {journal} {JHEP}\ }\textbf {\bibinfo {volume} {05}},\ \bibinfo {pages} {113} (\bibinfo {year} {2023})},\ \Eprint {http://arxiv.org/abs/2210.08021} {arXiv:2210.08021 [hep-ph]} \BibitemShut {NoStop}%
\bibitem [{\citenamefont {Acero}\ \emph {et~al.}(2022)\citenamefont {Acero} \emph {et~al.}}]{Acero:2022wqg}%
  \BibitemOpen
  \bibfield  {author} {\bibinfo {author} {\bibfnamefont {M.~A.}\ \bibnamefont {Acero}} \emph {et~al.},\ }\href@noop {} {\  (\bibinfo {year} {2022})},\ \Eprint {http://arxiv.org/abs/2203.07323} {arXiv:2203.07323 [hep-ex]} \BibitemShut {NoStop}%
\bibitem [{\citenamefont {Aguilar}\ \emph {et~al.}(2001)\citenamefont {Aguilar} \emph {et~al.}}]{LSND:2001aii}%
  \BibitemOpen
  \bibfield  {author} {\bibinfo {author} {\bibfnamefont {A.}~\bibnamefont {Aguilar}} \emph {et~al.} (\bibinfo {collaboration} {LSND}),\ }\href {\doibase 10.1103/PhysRevD.64.112007} {\bibfield  {journal} {\bibinfo  {journal} {Phys. Rev. D}\ }\textbf {\bibinfo {volume} {64}},\ \bibinfo {pages} {112007} (\bibinfo {year} {2001})},\ \Eprint {http://arxiv.org/abs/hep-ex/0104049} {arXiv:hep-ex/0104049} \BibitemShut {NoStop}%
\bibitem [{\citenamefont {Dentler}\ \emph {et~al.}(2018)\citenamefont {Dentler}, \citenamefont {Hern\'andez-Cabezudo}, \citenamefont {Kopp}, \citenamefont {Machado}, \citenamefont {Maltoni}, \citenamefont {Martinez-Soler},\ and\ \citenamefont {Schwetz}}]{Dentler:2018sju}%
  \BibitemOpen
  \bibfield  {author} {\bibinfo {author} {\bibfnamefont {M.}~\bibnamefont {Dentler}}, \bibinfo {author} {\bibfnamefont {A.}~\bibnamefont {Hern\'andez-Cabezudo}}, \bibinfo {author} {\bibfnamefont {J.}~\bibnamefont {Kopp}}, \bibinfo {author} {\bibfnamefont {P.~A.~N.}\ \bibnamefont {Machado}}, \bibinfo {author} {\bibfnamefont {M.}~\bibnamefont {Maltoni}}, \bibinfo {author} {\bibfnamefont {I.}~\bibnamefont {Martinez-Soler}}, \ and\ \bibinfo {author} {\bibfnamefont {T.}~\bibnamefont {Schwetz}},\ }\href {\doibase 10.1007/JHEP08(2018)010} {\bibfield  {journal} {\bibinfo  {journal} {JHEP}\ }\textbf {\bibinfo {volume} {08}},\ \bibinfo {pages} {010} (\bibinfo {year} {2018})},\ \Eprint {http://arxiv.org/abs/1803.10661} {arXiv:1803.10661 [hep-ph]} \BibitemShut {NoStop}%
\bibitem [{\citenamefont {Diaz}\ \emph {et~al.}(2020)\citenamefont {Diaz}, \citenamefont {Arg\"uelles}, \citenamefont {Collin}, \citenamefont {Conrad},\ and\ \citenamefont {Shaevitz}}]{Diaz:2019fwt}%
  \BibitemOpen
  \bibfield  {author} {\bibinfo {author} {\bibfnamefont {A.}~\bibnamefont {Diaz}}, \bibinfo {author} {\bibfnamefont {C.~A.}\ \bibnamefont {Arg\"uelles}}, \bibinfo {author} {\bibfnamefont {G.~H.}\ \bibnamefont {Collin}}, \bibinfo {author} {\bibfnamefont {J.~M.}\ \bibnamefont {Conrad}}, \ and\ \bibinfo {author} {\bibfnamefont {M.~H.}\ \bibnamefont {Shaevitz}},\ }\href {\doibase 10.1016/j.physrep.2020.08.005} {\bibfield  {journal} {\bibinfo  {journal} {Phys. Rept.}\ }\textbf {\bibinfo {volume} {884}},\ \bibinfo {pages} {1} (\bibinfo {year} {2020})},\ \Eprint {http://arxiv.org/abs/1906.00045} {arXiv:1906.00045 [hep-ex]} \BibitemShut {NoStop}%
\bibitem [{\citenamefont {B\"oser}\ \emph {et~al.}(2020)\citenamefont {B\"oser}, \citenamefont {Buck}, \citenamefont {Giunti}, \citenamefont {Lesgourgues}, \citenamefont {Ludhova}, \citenamefont {Mertens}, \citenamefont {Schukraft},\ and\ \citenamefont {Wurm}}]{Boser:2019rta}%
  \BibitemOpen
  \bibfield  {author} {\bibinfo {author} {\bibfnamefont {S.}~\bibnamefont {B\"oser}}, \bibinfo {author} {\bibfnamefont {C.}~\bibnamefont {Buck}}, \bibinfo {author} {\bibfnamefont {C.}~\bibnamefont {Giunti}}, \bibinfo {author} {\bibfnamefont {J.}~\bibnamefont {Lesgourgues}}, \bibinfo {author} {\bibfnamefont {L.}~\bibnamefont {Ludhova}}, \bibinfo {author} {\bibfnamefont {S.}~\bibnamefont {Mertens}}, \bibinfo {author} {\bibfnamefont {A.}~\bibnamefont {Schukraft}}, \ and\ \bibinfo {author} {\bibfnamefont {M.}~\bibnamefont {Wurm}},\ }\href {\doibase 10.1016/j.ppnp.2019.103736} {\bibfield  {journal} {\bibinfo  {journal} {Prog. Part. Nucl. Phys.}\ }\textbf {\bibinfo {volume} {111}},\ \bibinfo {pages} {103736} (\bibinfo {year} {2020})},\ \Eprint {http://arxiv.org/abs/1906.01739} {arXiv:1906.01739 [hep-ex]} \BibitemShut {NoStop}%
\bibitem [{\citenamefont {Acciarri}\ \emph {et~al.}(2015)\citenamefont {Acciarri} \emph {et~al.}}]{MicroBooNE:2015bmn}%
  \BibitemOpen
  \bibfield  {author} {\bibinfo {author} {\bibfnamefont {R.}~\bibnamefont {Acciarri}} \emph {et~al.} (\bibinfo {collaboration} {MicroBooNE, LAr1-ND, ICARUS-WA104}),\ }\href@noop {} {\  (\bibinfo {year} {2015})},\ \Eprint {http://arxiv.org/abs/1503.01520} {arXiv:1503.01520 [physics.ins-det]} \BibitemShut {NoStop}%
\bibitem [{\citenamefont {Abratenko}\ \emph {et~al.}(2022{\natexlab{a}})\citenamefont {Abratenko} \emph {et~al.}}]{MicroBooNE:2021zai}%
  \BibitemOpen
  \bibfield  {author} {\bibinfo {author} {\bibfnamefont {P.}~\bibnamefont {Abratenko}} \emph {et~al.} (\bibinfo {collaboration} {MicroBooNE}),\ }\href {\doibase 10.1103/PhysRevLett.128.111801} {\bibfield  {journal} {\bibinfo  {journal} {Phys. Rev. Lett.}\ }\textbf {\bibinfo {volume} {128}},\ \bibinfo {pages} {111801} (\bibinfo {year} {2022}{\natexlab{a}})},\ \Eprint {http://arxiv.org/abs/2110.00409} {arXiv:2110.00409 [hep-ex]} \BibitemShut {NoStop}%
\bibitem [{\citenamefont {Abratenko}\ \emph {et~al.}(2022{\natexlab{b}})\citenamefont {Abratenko} \emph {et~al.}}]{MicroBooNE:2021tya}%
  \BibitemOpen
  \bibfield  {author} {\bibinfo {author} {\bibfnamefont {P.}~\bibnamefont {Abratenko}} \emph {et~al.} (\bibinfo {collaboration} {MicroBooNE}),\ }\href {\doibase 10.1103/PhysRevLett.128.241801} {\bibfield  {journal} {\bibinfo  {journal} {Phys. Rev. Lett.}\ }\textbf {\bibinfo {volume} {128}},\ \bibinfo {pages} {241801} (\bibinfo {year} {2022}{\natexlab{b}})},\ \Eprint {http://arxiv.org/abs/2110.14054} {arXiv:2110.14054 [hep-ex]} \BibitemShut {NoStop}%
\bibitem [{\citenamefont {Abratenko}\ \emph {et~al.}(2022{\natexlab{c}})\citenamefont {Abratenko} \emph {et~al.}}]{MicroBooNE:2021nxr}%
  \BibitemOpen
  \bibfield  {author} {\bibinfo {author} {\bibfnamefont {P.}~\bibnamefont {Abratenko}} \emph {et~al.} (\bibinfo {collaboration} {MicroBooNE}),\ }\href {\doibase 10.1103/PhysRevD.105.112005} {\bibfield  {journal} {\bibinfo  {journal} {Phys. Rev. D}\ }\textbf {\bibinfo {volume} {105}},\ \bibinfo {pages} {112005} (\bibinfo {year} {2022}{\natexlab{c}})},\ \Eprint {http://arxiv.org/abs/2110.13978} {arXiv:2110.13978 [hep-ex]} \BibitemShut {NoStop}%
\bibitem [{\citenamefont {Abratenko}\ \emph {et~al.}(2022{\natexlab{d}})\citenamefont {Abratenko} \emph {et~al.}}]{MicroBooNE:2021pvo}%
  \BibitemOpen
  \bibfield  {author} {\bibinfo {author} {\bibfnamefont {P.}~\bibnamefont {Abratenko}} \emph {et~al.} (\bibinfo {collaboration} {MicroBooNE}),\ }\href {\doibase 10.1103/PhysRevD.105.112003} {\bibfield  {journal} {\bibinfo  {journal} {Phys. Rev. D}\ }\textbf {\bibinfo {volume} {105}},\ \bibinfo {pages} {112003} (\bibinfo {year} {2022}{\natexlab{d}})},\ \Eprint {http://arxiv.org/abs/2110.14080} {arXiv:2110.14080 [hep-ex]} \BibitemShut {NoStop}%
\bibitem [{\citenamefont {Abratenko}\ \emph {et~al.}(2022{\natexlab{e}})\citenamefont {Abratenko} \emph {et~al.}}]{MicroBooNE:2021wad}%
  \BibitemOpen
  \bibfield  {author} {\bibinfo {author} {\bibfnamefont {P.}~\bibnamefont {Abratenko}} \emph {et~al.} (\bibinfo {collaboration} {MicroBooNE}),\ }\href {\doibase 10.1103/PhysRevD.105.112004} {\bibfield  {journal} {\bibinfo  {journal} {Phys. Rev. D}\ }\textbf {\bibinfo {volume} {105}},\ \bibinfo {pages} {112004} (\bibinfo {year} {2022}{\natexlab{e}})},\ \Eprint {http://arxiv.org/abs/2110.14065} {arXiv:2110.14065 [hep-ex]} \BibitemShut {NoStop}%
\bibitem [{\citenamefont {Wang}\ \emph {et~al.}(2015)\citenamefont {Wang}, \citenamefont {Alvarez-Ruso},\ and\ \citenamefont {Nieves}}]{Wang:2014nat}%
  \BibitemOpen
  \bibfield  {author} {\bibinfo {author} {\bibfnamefont {E.}~\bibnamefont {Wang}}, \bibinfo {author} {\bibfnamefont {L.}~\bibnamefont {Alvarez-Ruso}}, \ and\ \bibinfo {author} {\bibfnamefont {J.}~\bibnamefont {Nieves}},\ }\href {\doibase 10.1016/j.physletb.2014.11.025} {\bibfield  {journal} {\bibinfo  {journal} {Phys. Lett. B}\ }\textbf {\bibinfo {volume} {740}},\ \bibinfo {pages} {16} (\bibinfo {year} {2015})},\ \Eprint {http://arxiv.org/abs/1407.6060} {arXiv:1407.6060 [hep-ph]} \BibitemShut {NoStop}%
\bibitem [{\citenamefont {Arg\"uelles}\ \emph {et~al.}(2022)\citenamefont {Arg\"uelles}, \citenamefont {Esteban}, \citenamefont {Hostert}, \citenamefont {Kelly}, \citenamefont {Kopp}, \citenamefont {Machado}, \citenamefont {Martinez-Soler},\ and\ \citenamefont {Perez-Gonzalez}}]{Arguelles:2021meu}%
  \BibitemOpen
  \bibfield  {author} {\bibinfo {author} {\bibfnamefont {C.~A.}\ \bibnamefont {Arg\"uelles}}, \bibinfo {author} {\bibfnamefont {I.}~\bibnamefont {Esteban}}, \bibinfo {author} {\bibfnamefont {M.}~\bibnamefont {Hostert}}, \bibinfo {author} {\bibfnamefont {K.~J.}\ \bibnamefont {Kelly}}, \bibinfo {author} {\bibfnamefont {J.}~\bibnamefont {Kopp}}, \bibinfo {author} {\bibfnamefont {P.~A.~N.}\ \bibnamefont {Machado}}, \bibinfo {author} {\bibfnamefont {I.}~\bibnamefont {Martinez-Soler}}, \ and\ \bibinfo {author} {\bibfnamefont {Y.~F.}\ \bibnamefont {Perez-Gonzalez}},\ }\href {\doibase 10.1103/PhysRevLett.128.241802} {\bibfield  {journal} {\bibinfo  {journal} {Phys. Rev. Lett.}\ }\textbf {\bibinfo {volume} {128}},\ \bibinfo {pages} {241802} (\bibinfo {year} {2022})},\ \Eprint {http://arxiv.org/abs/2111.10359} {arXiv:2111.10359 [hep-ph]} \BibitemShut {NoStop}%
\bibitem [{\citenamefont {Denton}(2022)}]{Denton:2021czb}%
  \BibitemOpen
  \bibfield  {author} {\bibinfo {author} {\bibfnamefont {P.~B.}\ \bibnamefont {Denton}},\ }\href {\doibase 10.1103/PhysRevLett.129.061801} {\bibfield  {journal} {\bibinfo  {journal} {Phys. Rev. Lett.}\ }\textbf {\bibinfo {volume} {129}},\ \bibinfo {pages} {061801} (\bibinfo {year} {2022})},\ \Eprint {http://arxiv.org/abs/2111.05793} {arXiv:2111.05793 [hep-ph]} \BibitemShut {NoStop}%
\bibitem [{\citenamefont {Abratenko}\ \emph {et~al.}(2023{\natexlab{a}})\citenamefont {Abratenko} \emph {et~al.}}]{MicroBooNE:2022sdp}%
  \BibitemOpen
  \bibfield  {author} {\bibinfo {author} {\bibfnamefont {P.}~\bibnamefont {Abratenko}} \emph {et~al.} (\bibinfo {collaboration} {MicroBooNE}),\ }\href {\doibase 10.1103/PhysRevLett.130.011801} {\bibfield  {journal} {\bibinfo  {journal} {Phys. Rev. Lett.}\ }\textbf {\bibinfo {volume} {130}},\ \bibinfo {pages} {011801} (\bibinfo {year} {2023}{\natexlab{a}})},\ \Eprint {http://arxiv.org/abs/2210.10216} {arXiv:2210.10216 [hep-ex]} \BibitemShut {NoStop}%
\bibitem [{\citenamefont {Aguilar-Arevalo}\ \emph {et~al.}(2021{\natexlab{b}})\citenamefont {Aguilar-Arevalo} \emph {et~al.}}]{MiniBooNE:2021bgc}%
  \BibitemOpen
  \bibfield  {author} {\bibinfo {author} {\bibfnamefont {A.~A.}\ \bibnamefont {Aguilar-Arevalo}} \emph {et~al.} (\bibinfo {collaboration} {MiniBooNE}),\ }\href@noop {} {\  (\bibinfo {year} {2021}{\natexlab{b}})},\ \Eprint {http://arxiv.org/abs/2110.15055} {arXiv:2110.15055 [hep-ex]} \BibitemShut {NoStop}%
\bibitem [{\citenamefont {Hostert}\ \emph {et~al.}(2024)\citenamefont {Hostert}, \citenamefont {Kelly},\ and\ \citenamefont {Zhou}}]{Hostert:2024etd}%
  \BibitemOpen
  \bibfield  {author} {\bibinfo {author} {\bibfnamefont {M.}~\bibnamefont {Hostert}}, \bibinfo {author} {\bibfnamefont {K.~J.}\ \bibnamefont {Kelly}}, \ and\ \bibinfo {author} {\bibfnamefont {T.}~\bibnamefont {Zhou}},\ }\href {\doibase 10.1103/PhysRevD.110.075002} {\bibfield  {journal} {\bibinfo  {journal} {Phys. Rev. D}\ }\textbf {\bibinfo {volume} {110}},\ \bibinfo {pages} {075002} (\bibinfo {year} {2024})},\ \Eprint {http://arxiv.org/abs/2406.04401} {arXiv:2406.04401 [hep-ph]} \BibitemShut {NoStop}%
\bibitem [{\citenamefont {Ashenfelter}\ \emph {et~al.}(2019)\citenamefont {Ashenfelter} \emph {et~al.}}]{PROSPECT:2018dnc}%
  \BibitemOpen
  \bibfield  {author} {\bibinfo {author} {\bibfnamefont {J.}~\bibnamefont {Ashenfelter}} \emph {et~al.} (\bibinfo {collaboration} {PROSPECT}),\ }\href {\doibase 10.1016/j.nima.2018.12.079} {\bibfield  {journal} {\bibinfo  {journal} {Nucl. Instrum. Meth. A}\ }\textbf {\bibinfo {volume} {922}},\ \bibinfo {pages} {287} (\bibinfo {year} {2019})},\ \Eprint {http://arxiv.org/abs/1808.00097} {arXiv:1808.00097 [physics.ins-det]} \BibitemShut {NoStop}%
\bibitem [{\citenamefont {Andriamirado}\ \emph {et~al.}(2021)\citenamefont {Andriamirado} \emph {et~al.}}]{PROSPECT:2020sxr}%
  \BibitemOpen
  \bibfield  {author} {\bibinfo {author} {\bibfnamefont {M.}~\bibnamefont {Andriamirado}} \emph {et~al.} (\bibinfo {collaboration} {PROSPECT}),\ }\href {\doibase 10.1103/PhysRevD.103.032001} {\bibfield  {journal} {\bibinfo  {journal} {Phys. Rev. D}\ }\textbf {\bibinfo {volume} {103}},\ \bibinfo {pages} {032001} (\bibinfo {year} {2021})},\ \Eprint {http://arxiv.org/abs/2006.11210} {arXiv:2006.11210 [hep-ex]} \BibitemShut {NoStop}%
\bibitem [{\citenamefont {Andriamirado}\ \emph {et~al.}(2023)\citenamefont {Andriamirado} \emph {et~al.}}]{PROSPECT:2022wlf}%
  \BibitemOpen
  \bibfield  {author} {\bibinfo {author} {\bibfnamefont {M.}~\bibnamefont {Andriamirado}} \emph {et~al.} (\bibinfo {collaboration} {PROSPECT, (PROSPECT Collaboration)*}),\ }\href {\doibase 10.1103/PhysRevLett.131.021802} {\bibfield  {journal} {\bibinfo  {journal} {Phys. Rev. Lett.}\ }\textbf {\bibinfo {volume} {131}},\ \bibinfo {pages} {021802} (\bibinfo {year} {2023})},\ \Eprint {http://arxiv.org/abs/2212.10669} {arXiv:2212.10669 [nucl-ex]} \BibitemShut {NoStop}%
\bibitem [{\citenamefont {Andriamirado}\ \emph {et~al.}(2024)\citenamefont {Andriamirado} \emph {et~al.}}]{PROSPECT:2024gps}%
  \BibitemOpen
  \bibfield  {author} {\bibinfo {author} {\bibfnamefont {M.}~\bibnamefont {Andriamirado}} \emph {et~al.} (\bibinfo {collaboration} {PROSPECT}),\ }\href@noop {} {\  (\bibinfo {year} {2024})},\ \Eprint {http://arxiv.org/abs/2406.10408} {arXiv:2406.10408 [hep-ex]} \BibitemShut {NoStop}%
\bibitem [{\citenamefont {{MicroBooNE Collaboration}}(2024{\natexlab{a}})}]{MicroBooNE-NOTE-1129-PUB}%
  \BibitemOpen
  \bibfield  {author} {\bibinfo {author} {\bibnamefont {{MicroBooNE Collaboration}}},\ }\href {https://microboone.fnal.gov/wp-content/uploads/MICROBOONE-NOTE-1129-PUB.pdf} {\emph {\bibinfo {title} {Updates to the NuMI Flux Simulation at MicroBooNE}}},\ \bibinfo {type} {Tech. Rep.}\ \bibinfo {number} {MICROBOONE-NOTE-1129-PUB}\ (\bibinfo  {institution} {MicroBooNE},\ \bibinfo {year} {2024})\BibitemShut {NoStop}%
\bibitem [{\citenamefont {{MicroBooNE Collaboration}}(2024{\natexlab{b}})}]{MicroBooNE-NOTE-1132-PUB}%
  \BibitemOpen
  \bibfield  {author} {\bibinfo {author} {\bibnamefont {{MicroBooNE Collaboration}}},\ }\href {https://microboone.fnal.gov/wp-content/uploads/MICROBOONE-NOTE-1132-PUB.pdf} {\emph {\bibinfo {title} {Search for a Sterile Neutrino in a 3+1 Framework using Wire-Cell Inclusive Charged-Current $\nu_e$ Selection with the BNB and NuMI beamlines in MicroBooNE}}},\ \bibinfo {type} {Tech. Rep.}\ \bibinfo {number} {MICROBOONE-NOTE-1132-PUB}\ (\bibinfo  {institution} {MicroBooNE},\ \bibinfo {year} {2024})\BibitemShut {NoStop}%
\bibitem [{\citenamefont {Adamson}\ \emph {et~al.}(2009)\citenamefont {Adamson} \emph {et~al.}}]{MiniBooNE:2008hnl}%
  \BibitemOpen
  \bibfield  {author} {\bibinfo {author} {\bibfnamefont {P.}~\bibnamefont {Adamson}} \emph {et~al.} (\bibinfo {collaboration} {MiniBooNE, MINOS}),\ }\href {\doibase 10.1103/PhysRevLett.102.211801} {\bibfield  {journal} {\bibinfo  {journal} {Phys. Rev. Lett.}\ }\textbf {\bibinfo {volume} {102}},\ \bibinfo {pages} {211801} (\bibinfo {year} {2009})},\ \Eprint {http://arxiv.org/abs/0809.2447} {arXiv:0809.2447 [hep-ex]} \BibitemShut {NoStop}%
\bibitem [{\citenamefont {Abratenko}\ \emph {et~al.}(2023{\natexlab{b}})\citenamefont {Abratenko} \emph {et~al.}}]{ICARUS:2023gpo}%
  \BibitemOpen
  \bibfield  {author} {\bibinfo {author} {\bibfnamefont {P.}~\bibnamefont {Abratenko}} \emph {et~al.} (\bibinfo {collaboration} {ICARUS}),\ }\href {\doibase 10.1140/epjc/s10052-023-11610-y} {\bibfield  {journal} {\bibinfo  {journal} {Eur. Phys. J. C}\ }\textbf {\bibinfo {volume} {83}},\ \bibinfo {pages} {467} (\bibinfo {year} {2023}{\natexlab{b}})},\ \Eprint {http://arxiv.org/abs/2301.08634} {arXiv:2301.08634 [hep-ex]} \BibitemShut {NoStop}%
\bibitem [{\citenamefont {Abratenko}\ \emph {et~al.}(2024)\citenamefont {Abratenko} \emph {et~al.}}]{SBND:2024vgn}%
  \BibitemOpen
  \bibfield  {author} {\bibinfo {author} {\bibfnamefont {P.}~\bibnamefont {Abratenko}} \emph {et~al.} (\bibinfo {collaboration} {SBND}),\ }\href {\doibase 10.1140/epjc/s10052-024-13306-3} {\bibfield  {journal} {\bibinfo  {journal} {Eur. Phys. J. C}\ }\textbf {\bibinfo {volume} {84}},\ \bibinfo {pages} {1046} (\bibinfo {year} {2024})},\ \Eprint {http://arxiv.org/abs/2406.07514} {arXiv:2406.07514 [physics.ins-det]} \BibitemShut {NoStop}%
\bibitem [{\citenamefont {Andriamirado}\ \emph {et~al.}(2022)\citenamefont {Andriamirado} \emph {et~al.}}]{PROSPECT:2021jey}%
  \BibitemOpen
  \bibfield  {author} {\bibinfo {author} {\bibfnamefont {M.}~\bibnamefont {Andriamirado}} \emph {et~al.} (\bibinfo {collaboration} {PROSPECT}),\ }\href {\doibase 10.1088/1361-6471/ac48a4} {\bibfield  {journal} {\bibinfo  {journal} {J. Phys. G}\ }\textbf {\bibinfo {volume} {49}},\ \bibinfo {pages} {070501} (\bibinfo {year} {2022})},\ \Eprint {http://arxiv.org/abs/2107.03934} {arXiv:2107.03934 [hep-ex]} \BibitemShut {NoStop}%
\bibitem [{\citenamefont {Goldhagen}\ \emph {et~al.}(2022)\citenamefont {Goldhagen}, \citenamefont {Maltoni}, \citenamefont {Reichard},\ and\ \citenamefont {Schwetz}}]{Goldhagen:2021kxe}%
  \BibitemOpen
  \bibfield  {author} {\bibinfo {author} {\bibfnamefont {K.}~\bibnamefont {Goldhagen}}, \bibinfo {author} {\bibfnamefont {M.}~\bibnamefont {Maltoni}}, \bibinfo {author} {\bibfnamefont {S.~E.}\ \bibnamefont {Reichard}}, \ and\ \bibinfo {author} {\bibfnamefont {T.}~\bibnamefont {Schwetz}},\ }\href {\doibase 10.1140/epjc/s10052-022-10052-2} {\bibfield  {journal} {\bibinfo  {journal} {Eur. Phys. J. C}\ }\textbf {\bibinfo {volume} {82}},\ \bibinfo {pages} {116} (\bibinfo {year} {2022})},\ \Eprint {http://arxiv.org/abs/2109.14898} {arXiv:2109.14898 [hep-ph]} \BibitemShut {NoStop}%
\bibitem [{\citenamefont {An}\ \emph {et~al.}(2014)\citenamefont {An} \emph {et~al.}}]{DayaBay:2014fct}%
  \BibitemOpen
  \bibfield  {author} {\bibinfo {author} {\bibfnamefont {F.~P.}\ \bibnamefont {An}} \emph {et~al.} (\bibinfo {collaboration} {Daya Bay}),\ }\href {\doibase 10.1103/PhysRevLett.113.141802} {\bibfield  {journal} {\bibinfo  {journal} {Phys. Rev. Lett.}\ }\textbf {\bibinfo {volume} {113}},\ \bibinfo {pages} {141802} (\bibinfo {year} {2014})},\ \Eprint {http://arxiv.org/abs/1407.7259} {arXiv:1407.7259 [hep-ex]} \BibitemShut {NoStop}%
\bibitem [{\citenamefont {Berryman}\ and\ \citenamefont {Huber}(2021)}]{Berryman:2020agd}%
  \BibitemOpen
  \bibfield  {author} {\bibinfo {author} {\bibfnamefont {J.~M.}\ \bibnamefont {Berryman}}\ and\ \bibinfo {author} {\bibfnamefont {P.}~\bibnamefont {Huber}},\ }\href {\doibase 10.1007/JHEP01(2021)167} {\bibfield  {journal} {\bibinfo  {journal} {JHEP}\ }\textbf {\bibinfo {volume} {01}},\ \bibinfo {pages} {167} (\bibinfo {year} {2021})},\ \Eprint {http://arxiv.org/abs/2005.01756} {arXiv:2005.01756 [hep-ph]} \BibitemShut {NoStop}%
\bibitem [{\citenamefont {Giunti}\ \emph {et~al.}(2022)\citenamefont {Giunti}, \citenamefont {Li}, \citenamefont {Ternes},\ and\ \citenamefont {Xin}}]{Giunti:2021kab}%
  \BibitemOpen
  \bibfield  {author} {\bibinfo {author} {\bibfnamefont {C.}~\bibnamefont {Giunti}}, \bibinfo {author} {\bibfnamefont {Y.~F.}\ \bibnamefont {Li}}, \bibinfo {author} {\bibfnamefont {C.~A.}\ \bibnamefont {Ternes}}, \ and\ \bibinfo {author} {\bibfnamefont {Z.}~\bibnamefont {Xin}},\ }\href {\doibase 10.1016/j.physletb.2022.137054} {\bibfield  {journal} {\bibinfo  {journal} {Phys. Lett. B}\ }\textbf {\bibinfo {volume} {829}},\ \bibinfo {pages} {137054} (\bibinfo {year} {2022})},\ \Eprint {http://arxiv.org/abs/2110.06820} {arXiv:2110.06820 [hep-ph]} \BibitemShut {NoStop}%
\bibitem [{\citenamefont {Cianci}\ \emph {et~al.}(2017)\citenamefont {Cianci}, \citenamefont {Furmanski}, \citenamefont {Karagiorgi},\ and\ \citenamefont {Ross-Lonergan}}]{Cianci:2017okw}%
  \BibitemOpen
  \bibfield  {author} {\bibinfo {author} {\bibfnamefont {D.}~\bibnamefont {Cianci}}, \bibinfo {author} {\bibfnamefont {A.}~\bibnamefont {Furmanski}}, \bibinfo {author} {\bibfnamefont {G.}~\bibnamefont {Karagiorgi}}, \ and\ \bibinfo {author} {\bibfnamefont {M.}~\bibnamefont {Ross-Lonergan}},\ }\href {\doibase 10.1103/PhysRevD.96.055001} {\bibfield  {journal} {\bibinfo  {journal} {Phys. Rev. D}\ }\textbf {\bibinfo {volume} {96}},\ \bibinfo {pages} {055001} (\bibinfo {year} {2017})},\ \Eprint {http://arxiv.org/abs/1702.01758} {arXiv:1702.01758 [hep-ph]} \BibitemShut {NoStop}%
\bibitem [{\citenamefont {Machado}\ \emph {et~al.}(2019)\citenamefont {Machado}, \citenamefont {Palamara},\ and\ \citenamefont {Schmitz}}]{Machado:2019oxb}%
  \BibitemOpen
  \bibfield  {author} {\bibinfo {author} {\bibfnamefont {P.~A.}\ \bibnamefont {Machado}}, \bibinfo {author} {\bibfnamefont {O.}~\bibnamefont {Palamara}}, \ and\ \bibinfo {author} {\bibfnamefont {D.~W.}\ \bibnamefont {Schmitz}},\ }\href {\doibase 10.1146/annurev-nucl-101917-020949} {\bibfield  {journal} {\bibinfo  {journal} {Ann. Rev. Nucl. Part. Sci.}\ }\textbf {\bibinfo {volume} {69}},\ \bibinfo {pages} {363} (\bibinfo {year} {2019})},\ \Eprint {http://arxiv.org/abs/1903.04608} {arXiv:1903.04608 [hep-ex]} \BibitemShut {NoStop}%
\end{thebibliography}%

\end{document}